\documentclass[12pt,preprint]{aastex}

\newcommand{\beq}{\begin{equation}}
\newcommand{\eeq}{\end{equation}}
\newcommand{\lsim}{\lesssim}
\newcommand{\gsim}{\gtrsim}

\begin{document}

\title{Actively Star Forming Elliptical Galaxies at Low Redshifts
in the Sloan Digital Sky Survey}

\author{ Masataka Fukugita\altaffilmark{1}, Osamu Nakamura\altaffilmark{1},
Edwin L. Turner\altaffilmark{2}, Joe Helmboldt\altaffilmark{3}, 
and R. C. Nichol\altaffilmark{4}}
\altaffiltext{1}{Institute for Cosmic Ray Research, University of Tokyo,
Kashiwa 277 8582, Japan; fukugita@icrr.u-tokyo.ac.jp}

\altaffiltext{2}{Princeton University Observatory,
Princeton, NJ08544, U. S. A.}

\altaffiltext{3}{Department of Astronomy, New Mexico State University,
P. O. Box 30001, Las Cruces, NM88003, U. S. A.}

\altaffiltext{4}{Department of Physics, Carnegie Mellon University, 
Pittburgh, PA 15232. U. S. A.}

\begin{abstract}

We report discovery of actively star forming elliptical galaxies
in a morphologically classified sample of bright galaxies at a low redshift
obtained from the Sloan Digital Sky Survey. The emission lines of these
galaxies do not show the characteristics of active galactic nuclei, and thus
their strong H$\alpha$ emission is ascribed to star formation with
a rate nearly as high as that is seen in typical late spiral galaxies. This is taken
as evidence against the traditional view that all elliptical galaxies formed
early and now evolve only passively. The frequency of such star forming elliptical
galaxies is a few tenths of a percent in the sample, but increases
to 3\% if we include active S0 galaxies.
We may identify these galaxies as probable progenitors of
so-called E+A galaxies that show the strong Balmer absorption feature of A
stars superimposed on an old star population.
The approximate match of the abundance of active elliptical plus S0
galaxies with that of E+A galaxies indicates that the duration of such late
star formation episodes is of the order of $\gsim 1$ Gyr.
If we interpret these galaxies as new additions to the early-type galaxy
population, and if we assume a power law for their number evolution,
the abundance of early-type galaxies at $z=1$ is about 30\% less than
that at $z=0$.
\end{abstract}

\section{Introduction}

Elliptical galaxy formation has been a key issue in the general theory
of the formation of galaxies.
It is the traditional view that elliptical galaxies are old
systems, which formed at a high redshift, passively evolving
until the present without further star formation activity
(Eggen, Lynden-Bell \& Sandage 1962; Tinsley \& Gunn 1976).
In recent decades a rival
view has been proposed based on the hierarchical structure formation scenario,
namely that
elliptical galaxies form from major mergers and are relatively
young objects (Kauffmann, White, Guiderdoni 1993; Baugh, Cole \&
Frenk 1996; see also Toomre 1977; Schweizer 2000). While many
observations support the old formation scenario for elliptical galaxies
at say, $z>2$ (Bower, Lucey \& Ellis 1992; Terlevich, Caldwell \& Bower 2001;
Ellis et al. 1997; Fukugita, Hogan \& Peebles 1996; Peebles 2001), there are
also recent reports indicating that field ellipticals
may have formed at rather low redshifts (van Dokkum \& Franx 2001;
True et al. 2002;
Menanteau, Abraham \& Ellis 2001).
Evidence has also been reported that a small fraction of early-type galaxies
retain a signature of star formation that occurred in the recent past
(Zabuldoff et al. 1996; Quintero et al. 2003; Goto et al. 2003).

In this Letter we present new evidence that a small fraction of
field elliptical galaxies show ongoing active
star formation with rates as high as those
in spiral galaxies.
The reason that such objects were not previously known may be due
to the fact that both a sufficiently
large spectroscopic elliptical galaxy
sample is needed, yet their morphology should be determined by laborious
by visual inspections.  In
other large-scale surveys, early-type galaxies are
selected by spectral features or by colour information, which
excludes actively star forming galaxies from the
sample (e.g., Bernardi et al. 2003; Eisenstein et al. 2003).
Actively star forming elliptical galaxies are here reported
in a bright galaxy sample, which was given morphologically classified
by visual inspections,
obtained from the Sloan Digital Sky Survey
(SDSS; York et al. 2000).

\section{The sample}

The SDSS contains both photometric (Gunn et al. 1998;
Hogg et al. 2001; Pier et al. 2003)
and spectroscopic (Blanton et al. 2003) surveys, and
produces the most homogeneous galaxy data set available to date.
The initial SDSS observations were made in the northern and
southern equatorial stripes, and produced a galaxy catalogue to $r'=22.5$
mag in five colour bands (Fukugita et al. 1996) with a photometric
calibration using standard stars observed at USNO (Smith et al. 2002).
Spectroscopic follow-up was made to 17.8 mag with accurately defined
criteria for target selection (Strauss et al. 2002).
The dominant part of these data has
already been published as an {\it Early Data Release} (EDR) (Stoughton et al.
2002) and {\it Data Release One} (Abazajian et al. 2003).  The sample is
particularly useful to study statistical properties of field galaxies,
which has traditionally been a difficult task.

The region of the sky we consider is the northern equatorial stripe
(SDSS photometry run numbers 752 and 756)
for $145.15^\circ\leq {\rm R.A.}\leq 235.97$ and $|\delta|\leq 1.27
^\circ$, which is included
in the EDR sample. The total area is 229.7 square deg.
We take bright galaxies with $r^*\leq 15.9$ mag after
Galactic extinction corrections. This magnitude limit corresponds
to the faintest galaxies for which reliable morphological classifications
can be made by eye using SDSS imagery for essentially all objects.
We have classified all galaxies satisfying this magnitude criterion
in the northern equatorial stripe. The total
number of galaxies is 1875, among which 1600 (85\%) also have spectroscopic
information. These galaxies have typical redshifts $z\lsim 0.12$.

All galaxies in our sample (1875)
are classified into one of seven morphological classes,
$T=0$ (corresponding to E in the Hubble type),
1 (S0), 2 (Sa), 3 (Sb), 4 (Sc), 5 (Sd), and 6 (Im).
Morphology classification was carried out by two of us
(MF and ON) using the $g'$ band image to enhance 
the galactic structure, according
to {\it Hubble Atlas of Galaxies} (Sandage 1961).
We assign an index of $-1$ when the images are so disturbed that
we cannot determine a morphological type.
The classifications by MF and ON agree within $\Delta T\leq 1.5$
for most galaxies and a mean of the two (0.5 step in $T$) is taken as the
final classification.

Among the 1600 galaxies
in the spectroscopic subsample we obtain 210 E galaxies, 251 S0 galaxies
and 169 galaxies for a class between E and S0, which may be
either true E/S0 objects or galaxies which were classified as E by one person
and S0 by the other.
The physical properties of our E/S0 galaxies are in fact
between the E and S0 samples as a whole.
We impose a rather strict criterion for the selection of E galaxies
that the galaxy shows no structure and is judged to be an ellitptical
galaxy by both clasifiers,
so that the sample does not contain contaminants from S0.
We obtain 926 spiral and 23 irregular galaxies.
21 galaxies are left unclassified.
The somewhat larger fraction of E-S0 galaxies (0.39) compared to
the value ($\approx 0.3$) usually
adopted in the literature (e.g., Fukugita, Hogan \& Peebles 1998)
is due to our use of $r'$ colour as the prime passband to make
the galaxy catalogue.

\section{Emission line galaxies in the elliptical galaxy sample}

Among 210 E galaxies, 3 galaxies show H$\alpha$ emission lines with
the equivalent widths greater than $10$\AA, which is a typical
value for star-forming spiral galaxies\footnote{There is one
more elliptical galaxy that shows emission feature 
(photometry id in {\it EDR}: 752-4-8-9469-0046). Since there is a close
companion to this galaxy, we remove it from our consideration.}. 
Among these three one
galaxy is considered to be a Seyfert 2 galaxy
due to strong [OIII]$\lambda5007$
emission. The other two either do not show [OIII] lines or
show only weak [OIII] features, and
are thus apparently actively star forming galaxies. This statement can be made
more quantitative by employing the line diagnostics in the diagram
of [OIII]/H$\beta$ versus [NII]/H$\alpha$ (Baldwin, Phillips, \&
Terlevich 1981; Velleuix \& Osterbrock 1987)
with the criterion defined by Kauffmann et al. (2003) for
the SDSS sample. We present the $g'$ band images (in the logartithmic sacle)
and the spectra
of the three galaxies in Figure 1 (the fourth panel shows a {\it bona fide}
elliptical galaxy for comparison). The contrast is adjusted to emphasize
the faintest level.  We cannot
detect any morphological differences among the four images. The
spectroscopic features are conspicuously different. The first two
show strong H$\alpha$ emission, whereas [OIII] features are missing
or weak. The spectra resemble those of late spiral or
irregular galaxies with active star formation. The star formation rates
are estimated to be $3-5M_\odot$ yr$^{-1}$ assuming the conventional
transformation formula of Glazebrook et al. (1999) for the Salpeter
initial mass function. The two galaxies have luminosities close to
the characteristic value ($M_{r^*}=21.4$ mag at $H_0=70$km~s$^{-1}$Mpc$^{-1}$),
so they are bona fide `giant elliptical galaxies'.
The third example shows strong H$\alpha$ and [OIII] emission, typical
of Seyfert 2 galaxies. The coordinates and the properties of these
three galaxies are given in Table 1.

The inverse concentration indicies as defined by the ratio of the radii
that contain 90\% and 50\% of the Petrosian flux, $c=r_{50}/r_{90}$,
are 0.39 and 0.33 for the two star forming galaxies. The former value
is near the upper edge of the $c$ parameter expected for elliptical
galaxies (Shimasaku et al. 2001):
see Figure 2, where the two galaxies are denoted by larger
solid circles at $T=0$. The colour of the two galaxies is significantly
bluer than that of elliptical colours (Figure 3a). Therefore, this type
of elliptical galaxy would be rejected in samples of early type
galaxies constructed on the basis of colours, as in e.g., Eisenstein et al. (2003)
and Bernardi et al. (2003).
We find that the two galaxies are slightly
bluer, by 
$\Delta (g^*-r^*)=(g^*-r^*)_{3''{\rm aperture}}-(g^*-r^*)_{\rm Petrosian} 
\simeq -0.1$ ,
near their centers (observed with
the $3''$ aperture) than the mean (using Petrosian magnitudes),
which is a reminiscent of with the blue core of higher redshift elliptical
galaxies found by
Menanteau et al. (2001). This contrasts to
normal early type galaxies which show the opposite colour gradient
$\Delta (g^*-r^*)\simeq +0.05$.  It is also opposite to normal
disc galaxies, for which the colour gradient is $\Delta (g^*-r^*)>0$.
The two star forming galaxies are in the general field, not in clusters.
The galaxy No. 2 (see Table 1) is in somewhat a rich environment in sky,
but none in the vicinity has a comparable redshift. 

As a check of the possibility that an error in the SDSS database was 
not responsible for the unusual association of spectra and morphology 
reported here, independent spectra of  these two galaxies
were obtained with the APO 3.5m telescope using the DIS 
instrument; the 3.5m data confirmed the presence of the strong emission 
lines in these objects. 
We conclude that 2 out of 210 elliptical galaxies show active star
formation. The
corresponding frequency is 2/1600=0.1\% in the entire sample of
galaxies. These two galaxies have cores comparable to or slightly
softer than ordinary elliptical galaxies.

We may extend this analysis to include E/S0 and S0 galaxies. We find that
19 out of 420 E/S0-S0 galaxies show H$\alpha$ lines with equivalent
widths of $>10$\AA.
This rate is significantly higher than that for elliptical galaxies alone.
Comparisons of the concentration index and the colour
are included in Figures 2 and 3 above. The distribution of the $c$ parameter
does not differ from that for normal S0 galaxies, but the $g^*-r^*$
colour is significantly bluer for most of emission line S0 galaxies.
The core is also bluer than the average colour by 0.05$-$0.1 mag.
Including E galaxies, the frequency of star forming early type galaxies
is 3.3\% in the early galaxy sample, and 1.3\% in the entire galaxy
sample at $z<0.12$.

If we include galaxies with Balmer emission of $>5$\AA~ equivalent width,
we add 4 galaxies in the E sample, and 4 galaxies in the E/S0-S0
sample. The fraction of early-type emission line galaxies in the
total galaxy sample then becomes 1.8\%.

\section{Discussion and Conclusions}

There is a class of galaxies, so-called E+A galaxies (Dressler \& Gunn 1983),
which show spectral features of A-stars, such as strong Balmer absorption,
superimposed on the old K-M star spectrum of early type galaxies. 
These galaxies are
believed to be those which are within 1 Gyr after star formation ceased.
There are a number of filed galaxies that show the E+A
feature (Zabludoff et al. 1996; Quintero et al. 2003; Goto et al. 2003).
We have plotted
{\it early-type galaxies} in the equivalent widths
EW(H$\alpha$)-EW(H$\delta$) plane in Figure 4.
If we impose the criterion 
EW(H$\delta)<-3$\AA~
(minus means absorption) we are left with 15 galaxies, among which one
(non AGN) S0 galaxy shows a strong H$\alpha$ emission with
EW(H$\alpha)>10$\AA.
We also see 
several galaxies which show H$\alpha$ emission and at the same time
significant H$\delta$ absorption. These galaxies may be identified as
transition cases in which star formation is currently declining. 
The unique feature of our analysis is that 
E+A galaxies are selected from the visually identified E and S0 galaxy sample.
This evidences the existence of E+A galaxies having early-type
morphology, whereas evidence has been reported that some
of E+A galaxies found in earlier literature are disc galaxies 
(e.g., Franx 1993; Caldwell et al. 1996).

The abundance of E+A galaxies depends on the selection criterion.
If we use more strict criterion, as in Zabludoff et al. 1996,
EW(H$\delta)<-5.5$\AA, the number decreases to 4. The fraction 0.25\%
in our sample is consistent with 21/11113=0.19\% ($0.05<z<0.13$)
derived by those authors
 from the LCRS sample. Quintero et al. (2003) identified 1200 E+A galaxies in
an SDSS sample of 156,000 galaxies (0.76\%). This fraction
is consistent with ours derived here, although the selection criteria
are different.

Our fraction of E+A galaxies, 0.9\%, (the number becomes 2.1\% if we take
EW(H$\delta)<-2$\AA~ as E+A galaxies)
is on the same order of magnitude as that of emission-line early-type galaxies,
although this comparison is admittedly qualitative. 

We would suppose that the emission-line  early-type
galaxies studied here develop into E+A galaxies after star formation ceases,
and further develop into normal early-type galaxies in a
few gigayears. With this assumption the relative frequencies of 
star forming vs. E+A galaxies imply that
early-type galaxies show star-formation activity for 1-2 Gyr, or at least
not much less than 1 Gyr. 
Since the lookback time corresponding to our maximum sight, $z\leq 0.12$,
1.5 Gyr is comparable to the duration of star-formation activity,
the star-formation activity once it took place below $z=0.12$
should be visible in our sample.

We do not ask the question how elliptical galaxies acquire star
formation activity.
If we would take the view of hierarchical galaxy formation that elliptical
galaxies formed by merging processes, and 
if we identify star-forming early-type galaxies as new
additions to the early-type galaxy population, the number of early-type
galaxies increases with time.
Assuming a power law for the number evolution $N\sim(1+z)^{-\gamma}$
the increase is at a rate of
$N^{-1}(dN/dz)\Delta z\sim -\gamma\Delta z$ during the interval
of $\Delta z$ at $z\approx 0$, and we then obtain $\gamma\approx 0.4$
for $\Delta z\approx 0.068$, the median redshift of our sample.
This implies that the number density of early-type galaxies at $z\approx 1$
is smaller than the present day value by 30\%.

The slow evolution we derived may be compared with that of  
Im et al. (2002), who concludeda moderate decline of early-type
galaxies in number by a few tens of percent at
$z\sim 1$ from their study of the luminosity function. 
The evolution of early-type galaxies we observed is
significantly lower than that obtained by
Kauffmann, Charlot \& White (1996), who
identified early-type galaxies by using a colour cut predicted in
a passively evolving galaxy model and found
the number density of early type galaxies at $z=1$ to be 1/2-1/3
the present value, which agrees with the evolution 
inferred in $\Omega=1$ CDM models. In a $\Lambda$ cosmology,
the evolution is slower at low redshifts; a 30\% reduction is
inferred (Kauffmann \& Charlot 1998), which is consistent with our
finding.

\vspace{10pt}

Funding for the creation and distribution of the SDSS Archive has been provided
by the Alfred P. Sloan Foundation, the Participating Institutions, the National
Aeronautics and Space Administration, the National Science Foundation, the US
Department of Energy, the Japanese Monbukagakusho, and the
Max-Planck-Gesellschaft. The SDSS Web site is http://www.sdss.org/.
The SDSS is managed by the Astrophysical Research Consortium (ARC) for the
Participating Institutions. The Participating Institutions are The University of
Chicago, Fermilab, the Institute for Advanced Study, the Japan Participation
Group, Johns Hopkins University, Los Alamos National Laboratory, the
Max-Planck-Institut f\"{u}r Astronomie, the Max-Planck-Institut
f\"{u}r Astrophysik, New Mexico State
University, the University of Pittsburgh, Princeton University, the US
Naval Observatory, and the University of Washington.
MF is supported in part by the Grant in Aid of the Ministry of Education.

\clearpage
\begin{table*}
\begin{center}
\caption{Elliptical galaxies with H$\alpha$ emission lines (with
equivalent width $>10$\AA).}
\begin{tabular}{lcccccccc}
\tableline \tableline
ID  & RA & dec & EW(H$\alpha$) & $r^*$ & $g^*-r^*$ & $M_{r^*}$ & redshift & SFR\\
    &    &      &  (\AA)  &       &        &      &     & ($M_\odot$yr$^{-1}$) \\
\tableline
galaxy \#1   & $12^h 08^m 23.5^s$ & $+0^\circ 06' 37''$  & 11.4  & 14.79 & 0.63 & $-$21.49 & 0.041  &  2.9  \\
galaxy \#2   & $11^h 23^m 27.0^s$ & $-0^\circ 42' 49''$ & 38.2  &  15.48 & 0.55  & $-$20.79 & 0.041  & 4.8    \\
galaxy \#3   & $10^h 0^m 08.4^s$ & $+0^\circ 59' 05''$  & 52.8  & 15.87  & 0.49  & $-$19.82 & 0.031 &  AGN  \\
\tableline
\end{tabular}
\end{center}
The luminosities are for $H_0=70$km~s$^{-1}$Mpc$^{-1}$.
For the star formation rate (SFR),
the conversion factor of Glazebrook et al. (1999) [model BC96(kl96)] is
assumed.
\end{table*}%

\clearpage

\begin{figure}
\includegraphics[angle=-90,scale=0.6]{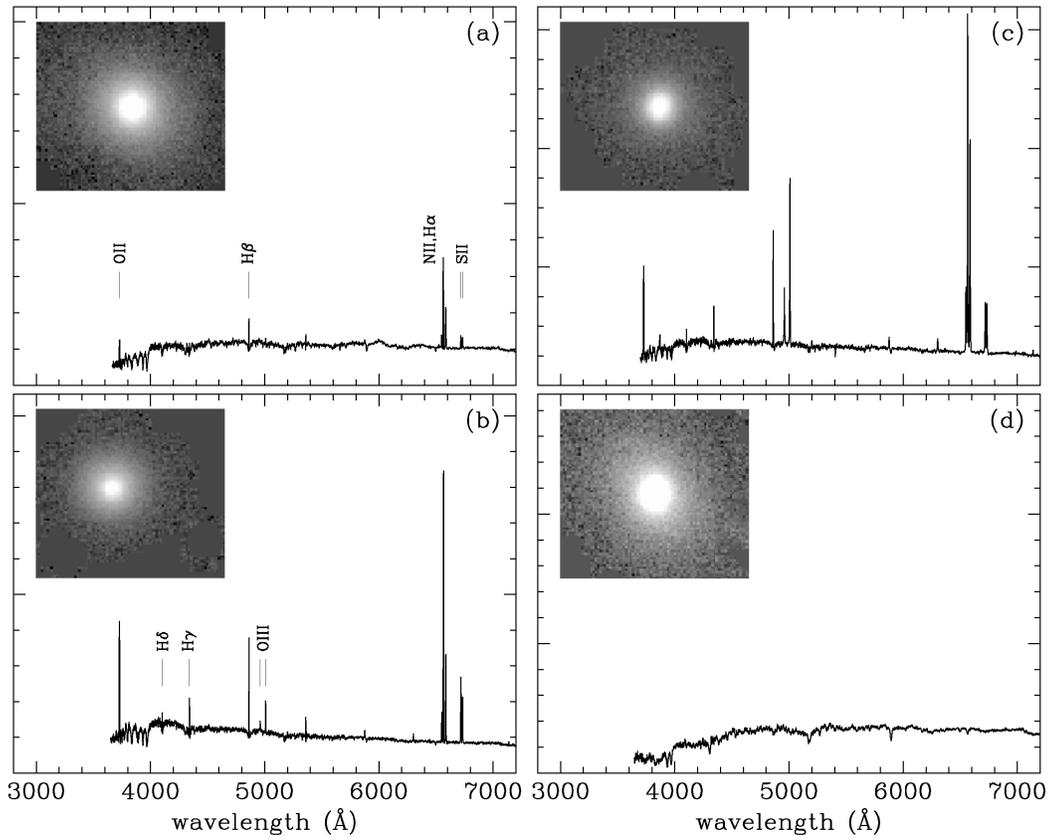}
\caption{
$r'$ band images and spectra. (a) and (b) elliptical
galaxies (galaxy \#1, and \#2) with strong H$\alpha$ emission lines,
where emission lines dominantly arise from star formation,
(c) H$\alpha$ emission-line elliptical galaxy with AGN activity
(galaxy \#3), and
(d) normal elliptical galaxy for comparison.
The identifications are those given in {\it EDR}.}
\end{figure}%

\begin{figure}
\includegraphics[angle=-90,scale=0.6]{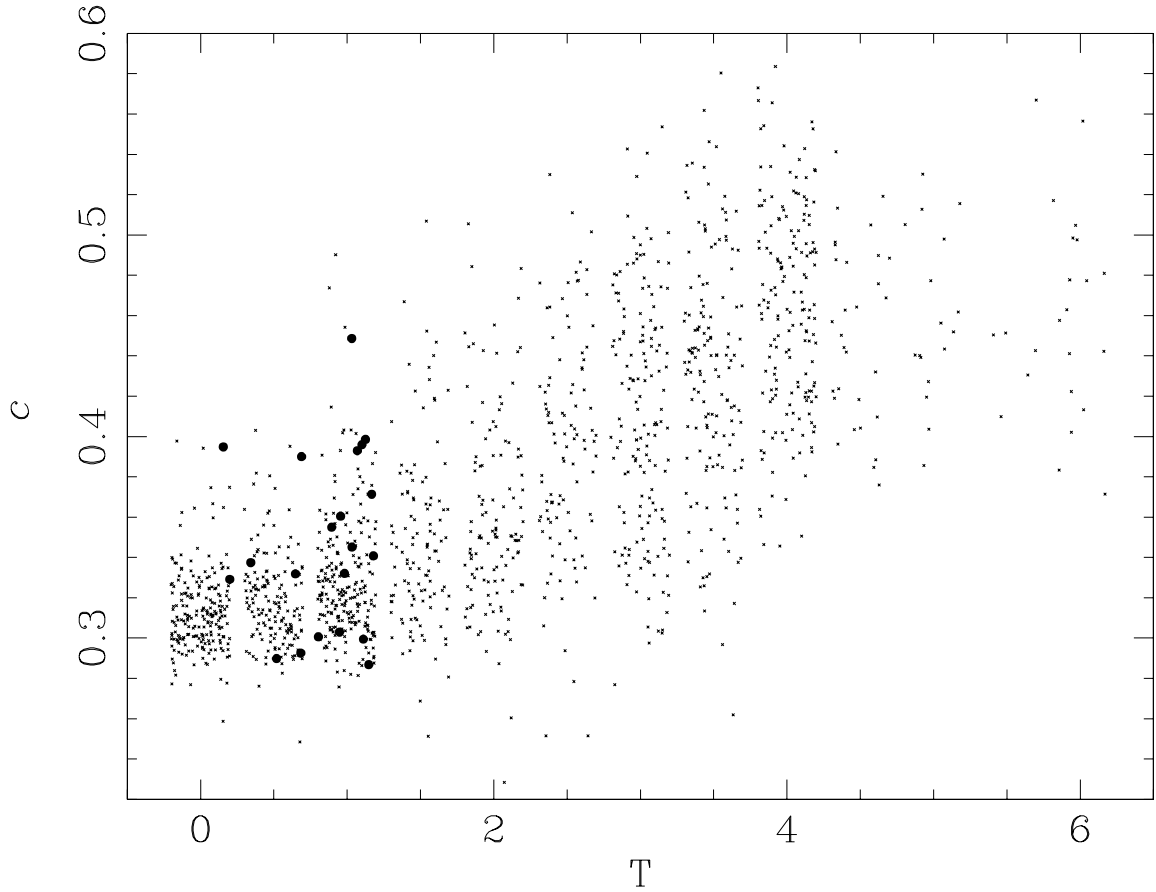}
\caption{
Inverse concentration parameters $C$ of morphologically classified
galaxies. Morphological classes are $T=0$ (E), 1 (S0), 2 (Sa), 3(Sb),
4(Sc), 5(Sd) and 6(Im) allowing for those in between (with $\Delta T=0.5$
step). Strong H$\alpha$ emission line galaxies of early types
are denoted by larger circles. We make the bins somewhat smeared to
avoid clutter of the data.}
\end{figure}%

\begin{figure}
\includegraphics[angle=-90,scale=0.6]{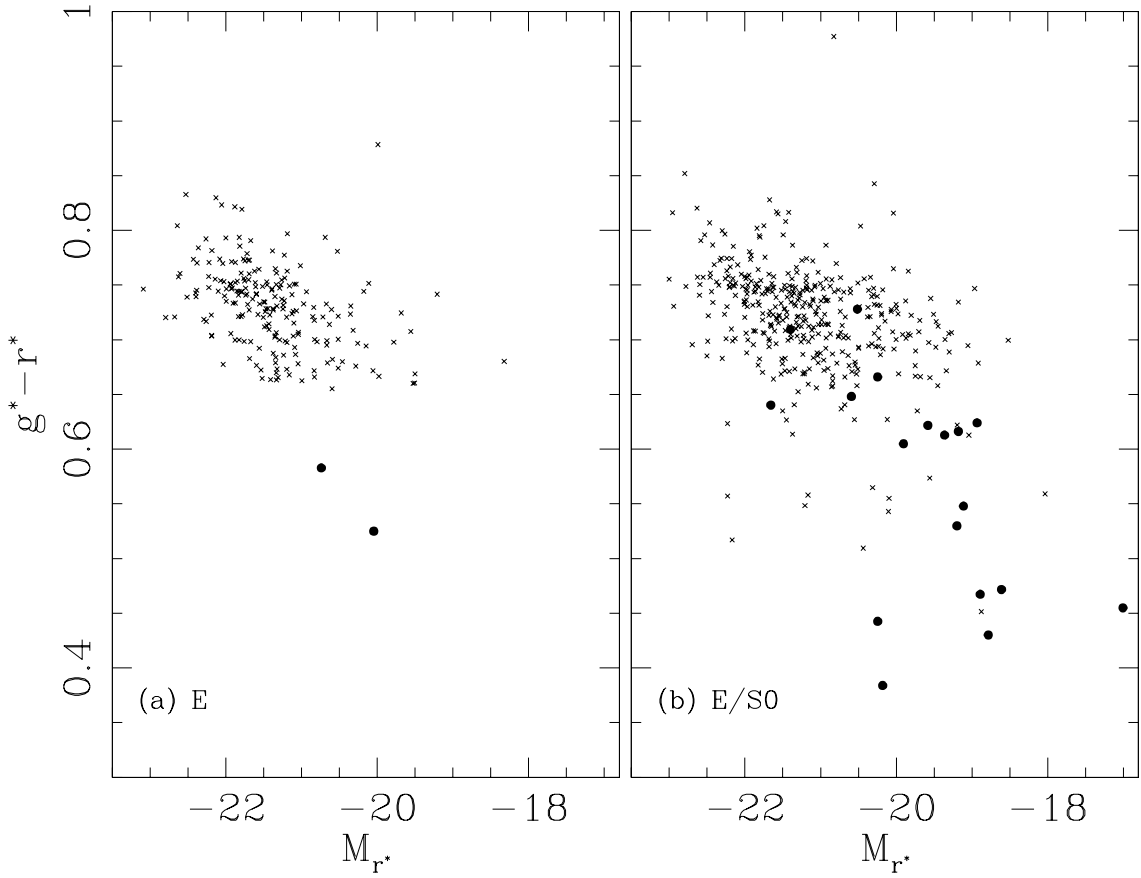}
\caption{
Colour magnitude diagram of early-type galaxies: (a) elliptical
galaxies, and (b) E/S0 and S0 galaxies. Galaxies that show strong H$\alpha$
emission are denoted by large circles. The subscript $g$ stands for the
Petrosian magnitude.}
\end{figure}%

\begin{figure}
\epsscale{1}
\plotone{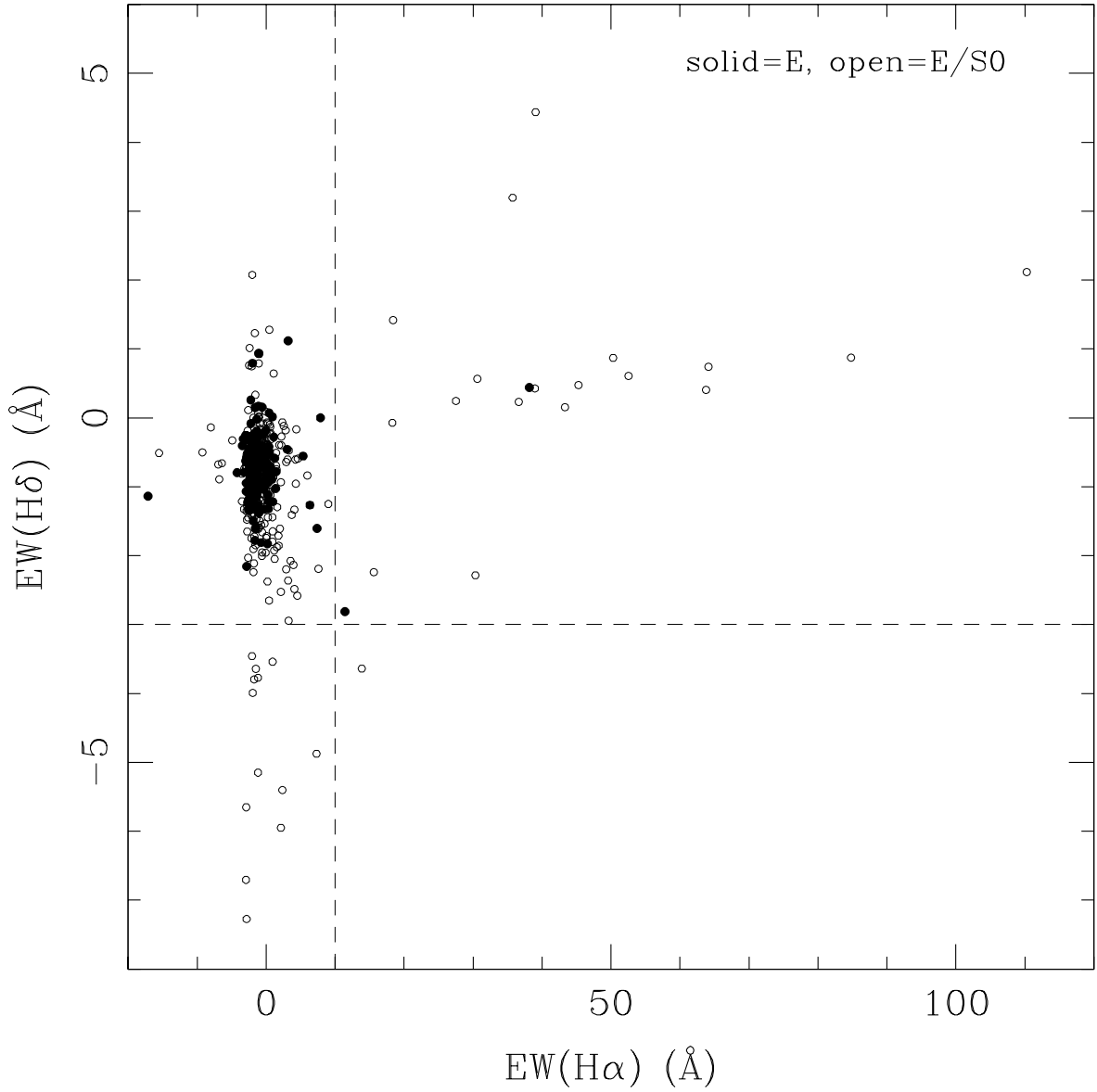}
\caption{
Distribution of early-type galaxies in the
EW(H$\alpha$)$-$EW(H$\delta$) plane. Solid circles denote
elliptical galaxies, and open circles are E/S0 to S0 galaxies.
Two dotted lines show the boundaries of
EW(H$\alpha$)$>10$\AA (strong emission-lines), and
EW(H$\delta$)$<-3$\AA (strong Balmer absorption features).}
\end{figure}%

\end{document}